\documentclass[twocolumn,showpacs,preprintnumbers,amsmath,amssymb,10pt,fleqn]{revtex4}

\usepackage{graphicx}% Include figure files
\usepackage{dcolumn}% Align table columns on decimal point
\usepackage{bm}% bold math
\usepackage{hyperref}
\input epsf

\begin{document}
\title{Possible observation sequences of Brans-Dicke wormholes}

\author{S.O. Alexeyev}
\email{alexeyev@sai.msu.ru}
\affiliation{%
  Sternberg Astronomical Institute, Lomonosov Moscow State University,
  Universitetsky Prospekt, 13, Moscow 119991, Russia}%
\author{K.A. Rannu}%
\email{melruin1986@gmail.com}
\affiliation{%
  Sternberg Astronomical Institute, Lomonosov Moscow State University,
  Universitetsky Prospekt, 13,Moscow 119991, Russia}%
\author{D.V. Gareeva}%
\email{4elesta@mail.ru}
\affiliation{%
  Faculty of Physics, Lomonosov Moscow State University, \\
  Leninskie Gory, Moscow 119991 Russia}%
\date{\today}

\begin{abstract}
The purpose of this study is to investigate observational
features of Brans-Dicke wormholes in a case if they exist in our
Universe. The energy flux from accretion onto a Brans-Dicke
wormhole and the so-called ``maximum impact parameter'' are
studied (the last one might allow to observe light sources through a
wormhole throat). The computed values were compared with the
corresponding ones for GR-wormholes and Schwarzschild black
holes. We shown that Brans-Dicke wormholes are
quasi-Schwarzschild objects and should differ from GR wormholes
by about one order of magnitude in the accretion energy flux.
\end{abstract}



\maketitle

\section{Introduction}

Different types of wormholes were extensively studied recently,
especially in the framework of extended gravitational models. A
lot of attention was paid to wormholes with massless scalar
fields \cite{ellis,bronn2}, in multidimentional theories
\cite{clement1,clement2,clement3,bhawal,dotti}, in brane-world
models \cite{anch,bronn3,cam,lobo1}, in semiclassical gravitation
\cite{garat}, and for different equations of state addressing the
Dark Matter (or energy) issues \cite{sushkov}. An important
question naturally arises: would it be possible to distinguish
between wormholes in different gravitational models using
observational data available in the near future with better precision? 
There are several methods of
compact objects' study nowadays. Novikov and Shatsky
\cite{shat1,shat2,shat3} have extracted distinctive features of
gravitational lensing of light passing through the wormhole
throat. It this work we focus on the possible observations of
accretion disks around Brans-Dicke wormholes. According to the
latest results, there are gas accretion disks or gas clouds
around almost all galactic nuclei at distances from $0.1$ to
$10^3$ pc \cite{urry}. It is possible to evaluate the mass of the
object through the gas flux in the clouds. The accretion speed
allows to establish the existence of a surface and hence to
determine the object's nature because black holes and wormholes
do not have any surface in the opposite to neutron stars that
have the one. In the articles by T. Harko, Z. Kovacs and F.S.N.
Lobo \cite{lobo2,lobo3} it was shown that fluxes of accretion
onto different types of wormholes in General Relativity (GR)
should differ by orders of magnitude. This approach therefore
sounds meaningful. Our work is devoted to analogous
considerations for Brans-Dicke wormholes and studies some of
their topological aspects as well. We consider a stationary model
of accretion disk \cite{lobo3}. Isotropic coordinates are used,
as in the original Brans-Dicke formalism. All the results are
written down in Planck units $c = G = \hbar = 1$.

\section{Brans-Dicke wormholes}

Brans-Dicke theory is a scalar-tensor gravitational model that
leads to GR when the coupling constant $| \omega | \to \infty$.
The action is:
\begin{gather}
  S = \frac{1}{16\pi} \int d^4x \ \sqrt{-g} \ (\phi R +
  \frac{\omega}{\phi} \ g^{\mu\nu} \phi_{,\mu} \phi_{,\nu} + L_{matter}),
\end{gather}
where $R$ is the Ricci scalar, $\phi$ is a scalar field,
$g_{\mu\nu}$ is the metric tensor and $L_{matter}$ is the
contribution of matter fields. The corresponding field equations
are
\begin{gather}\label{eq:02}
  \begin{split}
    R_{\mu\nu} - \frac{1}{2} \ g_{\mu\nu} R &= \frac{8\pi}{\phi} \
    T_{\mu\nu} + \frac{\omega}{\phi^2} \left(\phi_{,\mu} \phi_{,\nu} -
      \frac{1}{2} \ g_{\mu\nu} \phi^{,\sigma} \phi_{,\sigma} \right)
    \\
    & + \ \frac{1}{\phi}(\nabla_{\mu} \nabla_{\nu} \phi - g_{\mu\nu} \
    \Box \phi), \\
    \Box \phi &= \frac{8\pi T}{2 \omega + 3},
  \end{split}
\end{gather}
where $T = T^{\mu}_{\mu}$, $T_{\mu\nu}$ is the stress-energy
tensor and $\Box$ is d'Alembert operator.

There is a set of energy conditions in GR \cite{hock1} that
imposes limits on the stress-energy tensor $T_{\mu\nu}$
\cite{caroll}. To form a wormhole, one needs a matter that breaks
the null energy condition but still remains stable. As usual,
Jordan-Brans-Dicke theory allows gravity to influence matter via
the space-time metric tensor. But the matter itself can change
the metric both directly and via the additional scalar field.
Thus, the gravitational constant $G$ depends on the scalar field
which is variable in space and time. In the theory just the
scalar field of Brans-Dicke plays the role of matter. This model 
does not restrict other kinds of matter or dust to be added 
of course, it just can not be completely matterless.  As the
Einstein tensor breaks the null energy condition due to its
definition, the right part of expression (\ref{eq:02}) also
breaks that condition.

There are four static spherically-symmetric solutions in
Brans-Dicke theory \cite{brans} but only the first and the fourth
ones are independent. Scalar fields in Brans-Dicke model should
satisfy the equation \cite{bhadra}
\begin{gather}\label{eq:03}
  \phi = \phi_0 \left(1 + \frac{1}{\omega + 2} \ \frac{M}{r} \right)
\end{gather}
up to first order in $1/r$. $M$ is the asymptotic mass of the
wormhole at infinity. The forth solution breaks this condition so
we consider only the first Brans-Dicke class of solutions with
the metric
\begin{gather}\label{eq:04}
  \begin{split}
    ds^2 &= -\left( \cfrac{1 - 1/x}{1 + 1/x} \right)^{2l} dt^2 +
    \left( 1 + \cfrac{1}{x} \right)^4 \left( \frac{1 - 1/x}{1 + 1/x}
    \right)^{n}
 \\
    &\quad \times \ (d \rho^2 + \rho^2 d \Omega^2), \\
    \phi &= \phi_0 \left( \cfrac{1 - 1/x}{1 + 1/x} \right)^{p}.
  \end{split}
\end{gather}
Here $x = \rho/B$, $\quad l = 1 / \lambda$, $\quad n = (\lambda -
C - 1) / \lambda$, $\quad p = C / \lambda$, $\rho$ is the
isotropic radial coordinate, $\lambda, \ B, \ C$ and $\omega$ are
the constants related by \cite{brans}:
\begin{gather}\label{eq:05}
  \lambda = \sqrt{\cfrac{2\omega+3}{2\omega+4}}, \quad B = \cfrac{M}{2
    \phi_0} \ \sqrt{\cfrac{2 \omega + 4}{2 \omega + 3}}, \quad C = - \
  \frac{1}{\omega + 2}.
\end{gather}
At infinity, the metric asymptotically approaches the
Schwarzschild solution
\begin{gather*}
  ds^2 = \left( \cfrac{1 - \cfrac{r_g}{4 \rho}}{1 + \cfrac{r_g}{4
        \rho}} \right)^2 dt^2 - \left( 1 - \cfrac{r_g}{4 \rho}
  \right)^4 \left(  d {\rho}^2 + {\rho}^2 d \Omega^2 \right)
\end{gather*}
if $\omega \to \pm \ \infty$. The case $\omega \to + \ \infty$
corresponds to black holes and the one $\omega \to - \ \infty$
leads to wormholes. Thus $\lambda \to 1$, $C \to 0$ and $B \to
M/2$ at the infinity.

Numerical values for the expressions in Brans-Dicke theory depend
on the value of the coupling constant $\omega$. So it should be
possible to fix $\left| \omega \right|$ from observational data.
As it was shown by Agnese and La Camera, the discussed wormhole
is traversable if $\omega < - \ 2$ \cite{agness}. Thus the scalar
field itself plays the role of the required exotic matter. Using
Cassini experiments on PPN measuring, it is easy to find that
$\left| \omega \right| > 50000$ and only such values are
considered.

\section{Flux and topology. Results}

\begin{figure}[htp]
  $$
  \epsfxsize=8cm
  \epsfbox{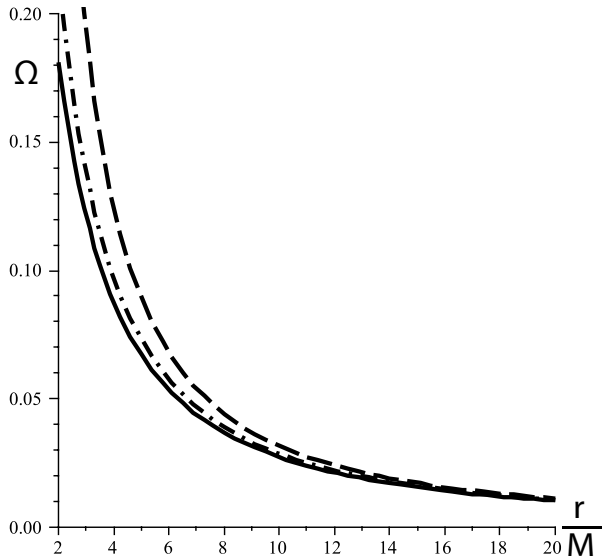}
  $$
\caption{Angular velocities of the accretion disk particles
($\times 10^{-4} \ \mbox{rad} / \mbox{c}$) for wormholes in GR
(dashdot line), Schwarzschild black hole (dash line) and
Brans-Dicke wormhole (solid line) as a function of the normalized
radius.}
\label{fig1}
\end{figure}

\begin{figure}[htp]
  $$
  \epsfxsize=8cm
  \epsfbox{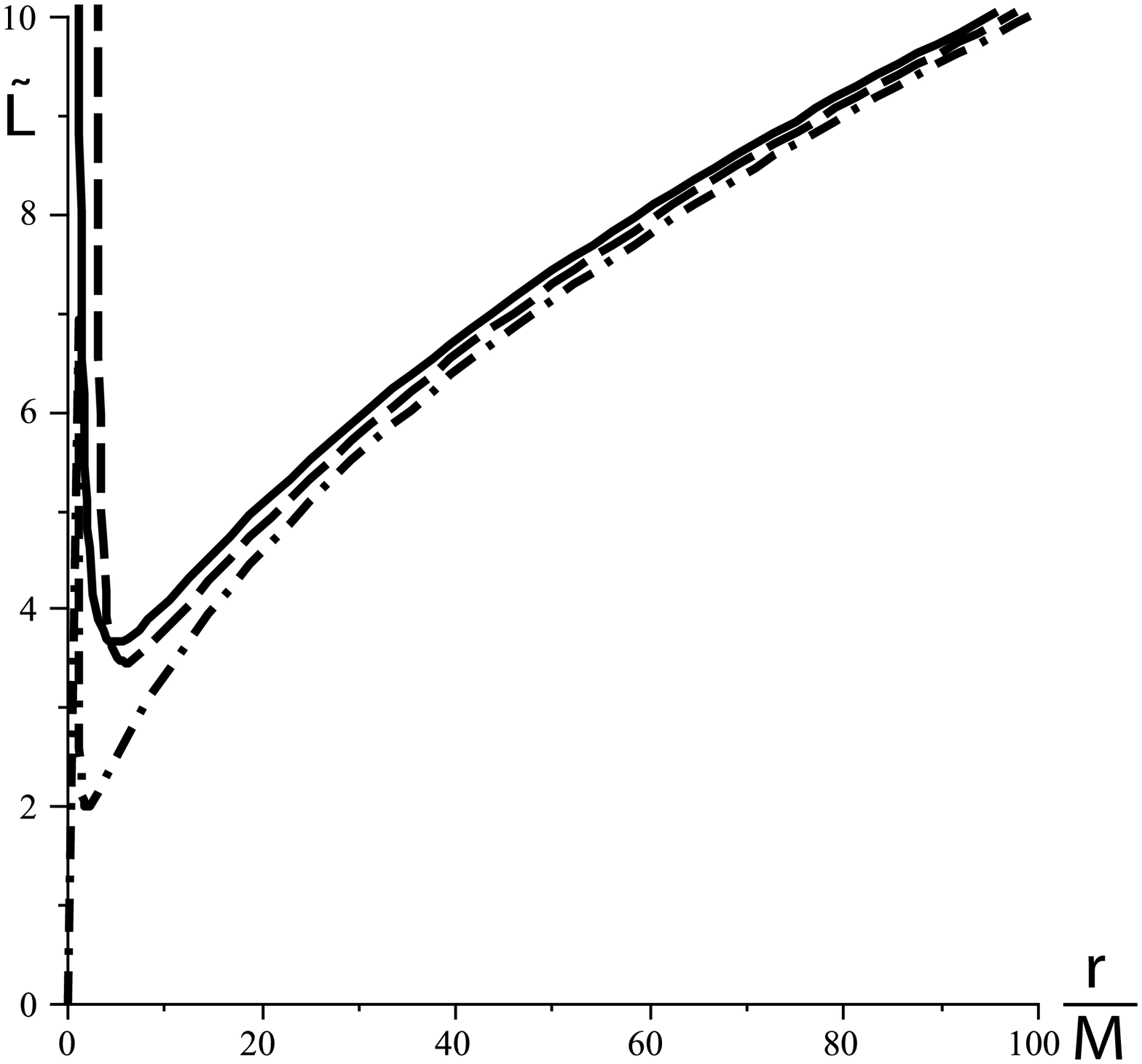}
  $$
\caption{Angular moments of the accretion disk particles ($\times
10^{4} \ \mbox{cm}^2 / \mbox{c}$) for wormholes in GR (dashdot
line), Schwarzschild black hole (dash line) and Brans-Dicke
wormhole (solid line) as a function the normalized radius.}
\label{fig2}
\end{figure}

\begin{figure}[htp]
  $$
  \epsfxsize=8cm
  \epsfbox{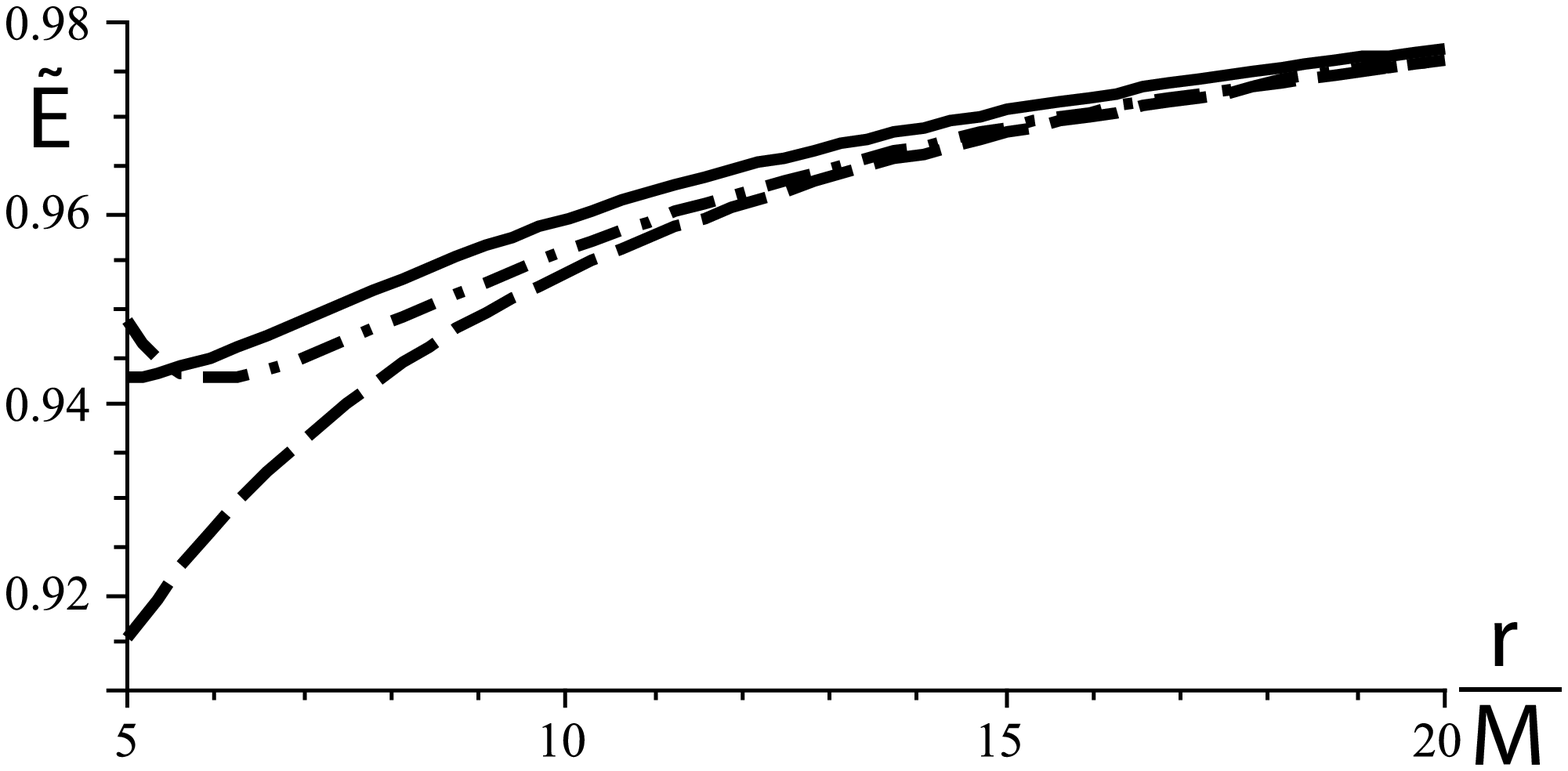}
  $$
\caption{Energy of the accretion disk particles ($\mbox{cm}^2 /
\mbox{c}^2$) in mass unites for wormholes in GR (dashdot line),
Schwarzschild black hole (dash line) and Brans-Dicke wormhole
(solid line) as a function of the normalized radius.}
\label{fig3}
\end{figure}

\begin{figure}[htp]
  $$
  \epsfxsize=8cm
  \epsfbox{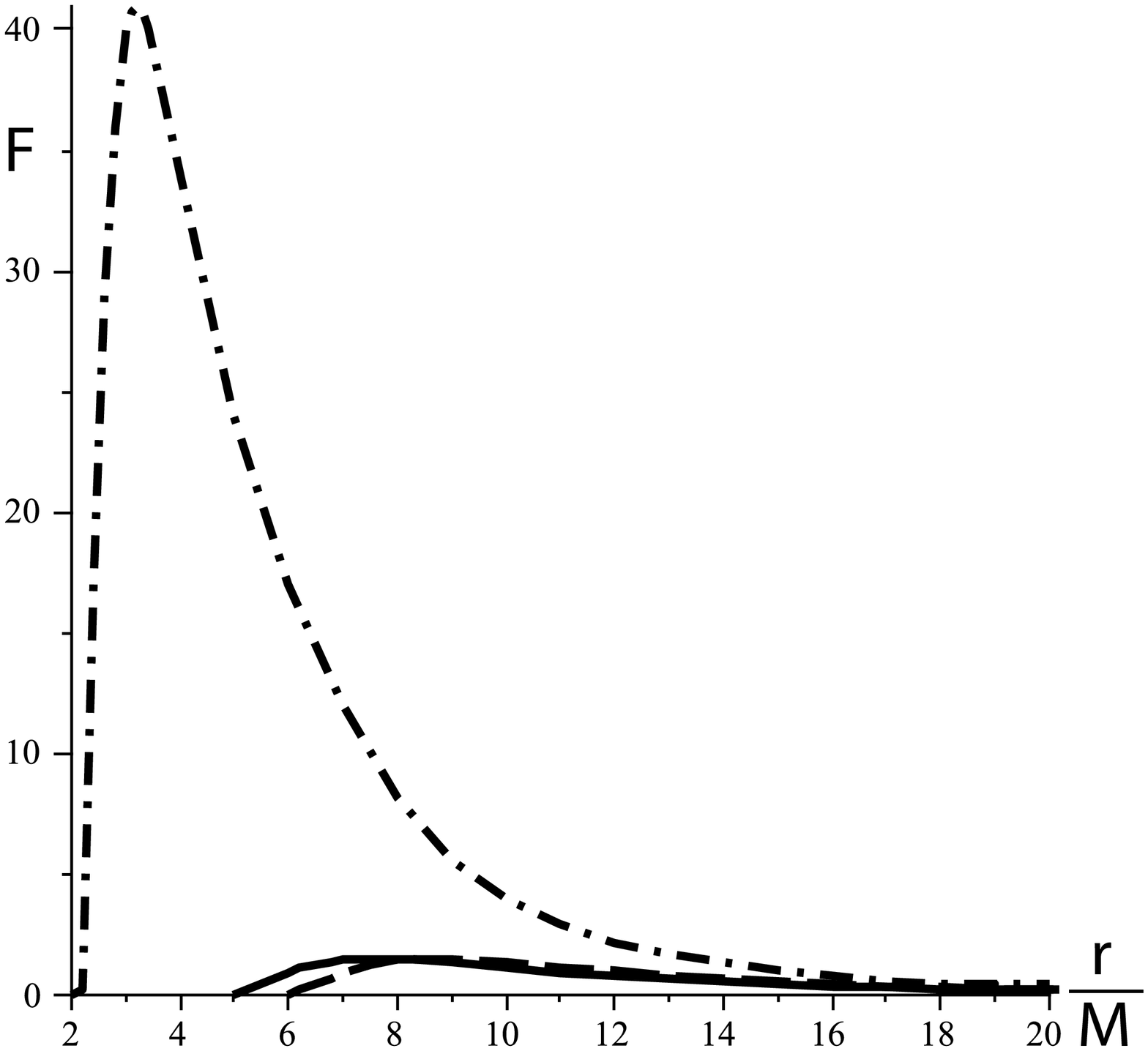}
  $$
\caption{Energy flux of the accretion disk particles ($\times
10^{23} \ \mbox{erg} / \mbox{cm}^2 / \mbox{c}$)for wormholes in
GR (dashdot line), Schwarzschild black hole (dash line) and
Brans-Dicke wormhole (solid line) as a function of the normalized
radius.}
\label{fig4}
\end{figure}

By solving geodesic equations, it is straightforward to establish
that energy, angular momentum and angular velocity for particles
on Keplerian orbits in the equatorial region of the accretion
disk are:
\begin{gather}\label{eq:06}
  \begin{split}
    \tilde{E} &= \left(\cfrac{x - \lambda}{x + \lambda} \right)^l
    \sqrt{\cfrac{x^2 + \lambda^2 - 2x \ (C + 1)}{x^2 + \lambda^2 - 2x
        \ (C + 2)}}, \\
    \tilde{L} &= \sqrt{\cfrac{2}{x}} \ B \ \cfrac{x^2 -
      \lambda^2}{\sqrt{x^2 + \lambda^2 - 2x \ (C + 2)}} \left(\cfrac{x
        + \lambda}{x - \lambda} \right)^{l+p}, \\
    \Omega &= \cfrac{x}{B} \ \cfrac{1}{x^2 - \lambda^2} \
    \sqrt{\cfrac{2x}{x^2 + \lambda^2 - 2x \ (C + 1)}} \left(\cfrac{x -
        \lambda}{x + \lambda} \right)^{p + 2l}.
  \end{split}
\end{gather}
The flux of energy emitted from the disk surface during the
accretion to the Brans-Dicke wormhole can be calculated
numerically after substituting (\ref{eq:06}) into the expression
\cite{pagethorn}:
\begin{gather}\label{eq:07}
  F(r) = - \ \frac{\dot{M_0}}{4 \pi \sqrt{-g}} \
  \frac{\Omega_{,r}}{(\tilde{E} - \Omega \tilde{L})^2} \int
  \limits_{r_{ms}}^{r} (\tilde{E} - \Omega \tilde{L}) \ \tilde{L}_{,r}
  \ dr,
\end{gather}
where $r_{ms}$ is the marginally stable orbit, and $\dot{M_0}$ is
the accretion speed $\cfrac{dM}{dt}$. On Fig.
\ref{fig1}--\ref{fig4} we compare the obtained values of angular
velocity and moment, energy and energy flux of the accretion disk
particles with the corresponding magnitudes for wormholes in GR
and Schwarzschild black holes.

\begin{figure}[htp]
  $$
  \epsfxsize=8cm
  \epsfbox{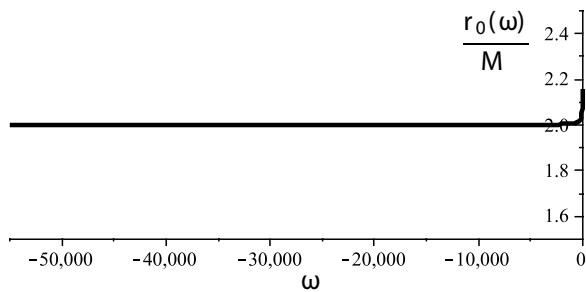}
  $$
\caption{Wormhole's throat radius in mass units as a function of
the Brans-Dicke parameter $\omega$.}
\label{fig5}
\end{figure}

The wormhole throat is the surface with the minimal possible area
surrounding the entry into another universe. The isotropic radial
coordinate $\rho$ on the throat surface is given by
\begin{gather}\label{eq:08}
  \rho_0 = \cfrac{\sqrt{2} B}{2} \left( \cfrac{2 \ | \omega + 1 | \pm
      \sqrt{- 8 - 6 \omega}}{\sqrt{(2 \omega + 3) (\omega + 2)}}
  \right).
\end{gather}
If $\left| \omega \right| > 50000$ the throat radius in arbitrary
coordinates is $r_0 = 2M$ (Fig. \ref{fig5}) with a high
precision. Hence, it coincides with the Schwarzschild
gravitational radius of the black hole with corresponding mass.

\begin{figure}[htp]
  $$
  \epsfxsize=8cm
 \epsfbox{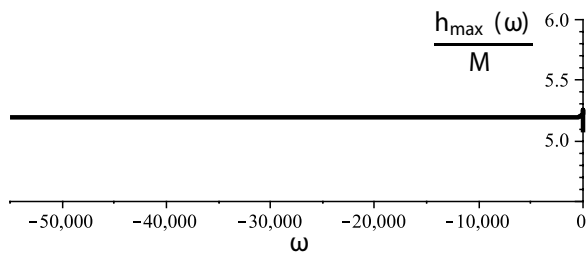}
  $$
\caption{Wormhole's maximum impact parameter in mass units as a
function of the Brans-Dicke parameter $\omega$.}
\label{fig6}
\end{figure}

The maximum impact parameter $h_{max}$ that allows to observe
light sources from the other universe \cite{shat1} for
Brans-Dicke wormholes in almost the full range of $\omega$ is
$h_{max} = 3 \sqrt{3} M\approx 5.18 M$ (Fig. \ref{fig6}). This
expected result confirms the fact that the observable Brans-Dicke
wormholes must be quasi-Schwarzschild (for them $\omega \to - \
\infty$ by definition).

The marginally stable orbit in the considered model was found
numerically and has a value $r_{ms} \approx 5M$.

One can find the stress-energy tensor for the Brans-Dicke scalar
field from:
 \begin{gather}\label{eq:09}
\begin{split}
      T_{\mu\nu} &= \frac{\omega}{\varphi^2} \left(\varphi_{,\mu} \varphi_{,\nu} -
        \frac{1}{2} \ g_{\mu\nu} \varphi^{,\sigma} \varphi_{,\sigma}
      \right) \\
        &+ \frac{1}{\varphi} \ (\nabla_{\mu} \nabla_{\nu} \varphi
      - g_{\mu\nu} \ \Box \varphi).
\end{split}
\end{gather}
Therefore
\begin{gather}\label{eq:10}
   \begin{split}
      T_{rr} &= \cfrac{\omega}{2} \ C^2 l^2
      \left(\cfrac{2}{\rho x \left(1 - 1/x^2 \right)} \right)^2 \left(
        \cfrac{1 - 1/x}{1 + 1/x} \right)^{2 C l} \\
      &+ \ \cfrac{1}{\rho^2} \left( \cfrac{1 - 1/x}{1 + 1/x}
      \right)^{C l} \left\{ 1 - \left( 1 + \cfrac{1}{x} \right)^4 \right. \\
        &\times \left. \left( \cfrac{1 - 1/x}{1 + 1/x} \right)^{2 \left( 1 - C l - l
          \right)} \right\} \\ 
       & \times \left[ \left( - \cfrac{2Cl}{x \left(1 -
              1/x^2 \right)} - 1 \right)^2 - \cfrac{4Cl}{x \left( 1 -
            1/x^2 \right)^2} - 1 \right].
    \end{split}
  \end{gather}
The study of this expression can reveal new properties of the
scalar field and will be the subject or further researches.

\section{Conclusions}

We have calculated the flux from accretion onto the Brans-Dicke
wormhole. It is important to underline that the observer at
infinity sees only the integral flux. The distribution of the
flux of energy emitted form the accretion disk is almost Gauss one,
so its integral flux is proportional to the maximum of the energy
one. These maxima are just the values to be compared. The maximum
energy for the accretion onto spherically-symmetric wormholes in
GR is one order of magnitude larger that the one for the
accretion onto the Brans-Dicke wormhole or on a Schwarzschild
black hole. As shown here, for the last two types the values or
the energy maximum are almost the same. As allowed values for
$\omega$ are already restricted, measuring of the energy flux
allows to test the considered model and can help to distinguish
between different types of compact objects in future.

The throat radius and maximum impact parameter for a Brans-Dicke
wormhole were found. We show that these values do not differ from
the ones associated with a quasi-Schwarzschild wormhole.
According to the Birkhoff theorem the Schwarzschild metric is the
most common spherically-symmetric one in a curved
space-time. This solution by itself describes the black hole and
does not lead to such objects as wormholes, but the studied
wormholes tend to the Schwarzschild metric at the infinity. So it
is possible to claim that Brans-Dicke wormholes are
asymptotically Schwarzschild. This fact allows to search for them
basing on future observational data with more accuracy.

\section{Acknowledgements}

Authors would like to thank Prof. Aurelien Barrau and Dr.
Alexander Shatsky for the useful discussions on the subject of
this work. The work was supported by the Federal Agency on
Science and Innovations of Russian Federation via State Contract
No. 02.740.11.0575.


\begin{thebibliography}{99}

\bibitem{ellis}
  H.G. Ellis,
  J. Math. Phys. {\bf 14}, 104 (1973)

\bibitem{bronn2}
  K.A. Bronnikov,
  Acta Phys. Polon. B {\bf 4}, 251 (1973) 

\bibitem{clement1}
  G. Clement,
  Gen. Rel. Grav. {\bf 16}, 131 (1984)

\bibitem{clement2}
  G. Clement,
  Gen. Rel. Grav. {\bf 16}, 477 (1984)

\bibitem{clement3}
  G. Clement,
  Gen. Rel. Grav. {\bf 16}, 491 (1984)

\bibitem{bhawal}
  B. Bhawal, S. Kar,
  Phys. Rev. D {\bf 46}, 2464 (1992)

\bibitem{dotti}
  G. Dotti, J. Oliva, R. Troncoso,
  Phys. Rev. D {\bf 75}, 024002 (2007)

\bibitem{anch}
  L.A. Anchordoqui, S.E. PerezBergliaffa,
  Phys. Rev. D {\bf 62}, 067502 (2000)

\bibitem{bronn3}
  K.A. Bronnikov, S.-W. Kim,
  Phys. Rev. D {\bf 67}, 064027 (2003)

\bibitem{cam}
  M. La Camera,
  Phys.Lett. B {\bf 573}, 27 (2003)

\bibitem{lobo1}
  F.S.N. Lobo,
  Phys. Rev. D {\bf 75}, 064027 (2007)

\bibitem{garat}
  R. Garattini, F.S.N. Lobo,
  Class. Quant. Grav. {\bf 24}, 2401 (2007)

\bibitem{sushkov}
  S. Sushkov,
  Phys. Rev. D {\bf 71}, 043520 (2005)

\bibitem{shat1}
  A.A. Shatsky,
  Phys.Usp. {\bf 52}, 811 (2009)

\bibitem{shat2}
  A.A. Shatsky, I.D. Novikov, N.S. Kardashev,
  Phys.Usp. {\bf 51}, 457 (2008)

\bibitem{shat3}
  I.D. Novikov, N.S. Kardashev, A.A. Shatsky,
  Phys.Usp. {\bf 50}, 965 (2007)

\bibitem{urry}
  C.M. Urry, P. Padovani,
  Publ. Astron. Soc. of the Pacific {\bf 107}, 803 (1995)

\bibitem{lobo2}
  T. Harko, Z. Kovacs, F.S.N. Lobo,
  Phys. Rev. D {\bf 78}, 084005 (2008)

\bibitem{lobo3}
  T. Harko, Z. Kovacs, F.S.N. Lobo,
  Phys. Rev. D {\bf 79}, 064001 (2009)

\bibitem{hock1}
  S.W. Hawking and G.F.R. Ellis,
  The Large Scale Structure of Space-Time, (Cambridge, Cambridge
  University Press: 1973)

\bibitem{caroll}
 S.M. Carroll, M. Hoffman, M. Trodden,
  Phys. Rev. D {\bf 68}, 023509 (2003)

\bibitem{brans}
  C.H. Brans, R.H. Dicke,
  Phys. Rev. {\bf 124}, 925 (1961)

\bibitem{bhadra}
  A. Bhadra, K. Sarkar,
  Mod. Phys. Lett. A {\bf 20}, 1831 (2005)

\bibitem{agness}
  A.G. Agnese, M. La Camera,
  Phys. Rev. D {\bf 51}, 2011 (1995)

\bibitem{pagethorn}
  D.N. Page, K.S. Thorn,
  Astrophys. J. {\bf 191}, 499 (1974)

\end{thebibliography}
\end{document}